\documentclass[12pt]{article}
\usepackage{amssymb,amsmath,epsfig}
\allowdisplaybreaks
\begin{document}
\title{\bf Stellar Evolution of Compact Stars in Curvature-Matter Coupling Gravity}

\author{M. Sharif \thanks{msharif.math@pu.edu.pk} and Arfa Waseem
\thanks{arfawaseem.pu@gmail.com}\\
Department of Mathematics, University of the Punjab,\\
Quaid-e-Azam Campus, Lahore-54590, Pakistan.}
\date{}

\maketitle

\begin{abstract}
This paper is devoted to study the stellar evolution of compact
objects whose energy density and pressure of the fluid are
interlinked by means of MIT bag model and a realistic polytropic
equation of state in the scenario of $f(R,T,Q)$ gravity, where
$Q=R_{ab}T^{ab}$. We derive the field equations as well as the
hydrostatic equilibrium equation and analyze their solutions
numerically for $R+\delta Q$ functional form with $\delta$ being a
coupling parameter. We discuss the dependence of various physical
properties such as pressure, energy density, total mass and surface
redshift on the chosen values of the model parameter. The physical
acceptability of the proposed model is examined by checking the
validity of energy conditions, causality condition, and adiabatic
index. We also study the effects arising due to matter-curvature
coupling on the compact stellar system. It is found that maximum
mass point lies within the observational range which indicates that
our model is appropriate to describe dense stellar objects.
\end{abstract}

\section{Introduction}

Gravitational collapse is a fascinating phenomenon, responsibles to
produce remnants of massive stars known as compact objects. Compact
objects are characterized by black holes, neutron stars and white
dwarfs depending upon the mass of stellar objects. The final
consequence of gravitational collapse entirely depends upon the
original mass of the celestial bodies. Stars having a mass less than
8 solar mass $(M_{\odot})$ construct white dwarfs whereas the stars
possessing larger mass, transform into neutron stars or black holes
as a result of the collapse. The existence of white dwarfs as well
as neutron stars can be detected in our universe but the presence of
black holes is affirmed only through some hypothetical observations.

The compact relativistic objects are extremely dense as well as
relatively small whose pair can combine together to generate
gravitational waves. Hawking \cite{4} observed an upper limit for
the gravitational radioactive energy ejected as a result of
concussion between black holes. Wagoner \cite{5} examined the energy
produced by the rotation of neutron stars. In a region of higher
gravitational potential, the observed electromagnetic radiations are
redshifted in frequency and this phenomenon is referred as surface
redshift $(z_{s})$ which describes the relation between the interior
geometry of star and its equation of state (EoS). The maximum bound
for the surface redshift corresponding to isotropic as well as
anisotropic spherical configurations are obtained as $z_{s}\leq2$
\cite{6} and $2\leq z_{s}\leq5$ \cite{7}, respectively. B\"{o}hmer
and Harko \cite{8} estimated the lower and upper limits of some
viable constituents in the context of anisotropic fluid
configuration with cosmological constant ($\Lambda$). They found
total energy bounds as well as redshift in terms of the anisotropic
factor.

It is well-known that modified theories of gravity help to
understand the current rapid expansion of cosmos. Such theories are
formulated by the modification of gravitational part in
Einstein-Hilbert action. The $f(R)$ gravity \cite{9} is the most
smooth modified form of general relativity (GR) proposed by
considering a generic form $f(R)$ instead of Ricci scalar in the
action of GR. Harko et al. \cite{10} developed $f(R, T)$ gravity by
including the matter effects in $f(R)$ theory, where T reveals the
trace component of the stress tensor. The coupling between matter
and gravitational parts yields a basic term which can produce
fruitful results. It may provide a matter based deviation from the
equation of motion and also assists to analyze dark source effects
as well as late-time acceleration. Motivated by this argument, a
more complicated and extended theory having strong non-minimal
curvature-matter combination is developed called $f(R,T,Q)$ gravity
\cite{11,12}.

It is observed that theories similar to $f(R,T)$ gravity are not the
most general Lagrangians describing the non-minimal coupling between
matter and geometry. An interesting difference in $f(R,T)$ gravity
and $f(R,T,Q)$ gravity is that for $T = 0$, the $f(R,T)$ field
equations reduce to $f(R)$ theory whereas in $f(R,T,Q)$ gravity, the
presence of $Q$ still contains the strong effects of non-minimal
coupling to the electromagnetic field. Sharif and Zubair
\cite{13,14} checked the consistency of energy conditions in this
theory and also discussed the validity of thermodynamical laws in
the same scenario. Ayuso et al. \cite{15} explored the physical
viability and stability of this theory. Yousaf et al. \cite{16}
discussed the anisotropic stable structure of the cylindrical system
as well as non-static spherical stellar models. In the same theory,
Sharif and Waseem \cite{17} investigated the regions of stable
Einstein universe with linear EoS by considering homogeneous and
inhomogeneous linear perturbations.

Compact objects are expressed as the basic ingredients in the field
of astrophysics and cosmology. The study of physical properties and
stability of these stars captivated the attention of many
researchers. Mak and Harko \cite{18} presented exact analytical
solutions using linear EoS which describe anisotropic static
spherical quark matter configuration. The same authors \cite{19}
also analyzed physical parameters (mass and radius) to discuss the
features of neutron stars in GR. Chaisi and Maharaj \cite{20}
studied the anisotropic distribution of compact stars with the
limiting case of energy density $\rho\propto r^{-2}$. Hossein et al.
\cite{21} investigated the structure of stellar system along with
variable $\Lambda$ which acts as a competent candidate of dark
energy. The power series solutions using barotropic and polytropic
EoS for perfect fluid distribution are also examined to analyze the
hydrostatic stability of dense stellar objects \cite{22}. Sharif and
Sadiq \cite{23} discussed the electromagnetic effects on the
stability of stars for two polytropic EoS using perturbations on
matter variables.

In recent years, the stability of self-gravitating system has become
a captivating subject in modified or alternative theories of
gravity. Sharif and Yousaf \cite{24} observed the viability of
spherical stellar system through perturbation approach in $f(R,T)$
theory. Abbas et al. \cite{25} established the physical
characteristics of particular strange quintessence star models in
$f(R)$ scenario. The stellar stable configuration of quark as well
as polytropic stars are also studied in $f(R,T)$ gravity \cite{26}.
The physical features of some specific star models for different
fluid distributions are investigated in $f(R,T,Q)$ gravity
\cite{27}. Deb et al. \cite{28} presented spherically symmetric
isotropic quark stars governed by the MIT bag model in $f(R,T)$
theory. They examined several physical properties of quark stars and
showed their behavior graphically. Sharif and Siddiqa \cite{28a}
analyzed the stellar models described by two different cases of
polytopic EoS with anisotropic distribution in $f(R,T)$ background.
Recently, we have discussed the behavior of anisotropic quark stars
by considering MIT bag model in the same framework. We have also
determined the graphical behavior of particular compact star
candidates \cite{28b}.

From an astrophysical point of view, the scenario of stellar system
investigates how stellar solution satisfies some general physical
requirements. This paper is therefore dedicated to examine the
physical conduct as well as the stability of compact stars to
determine the constraints for which the system of stellar equations
is physically realistic in $f(R,T,Q)$ gravity. The format of the
paper is presented as follows. In the next section, we present basic
formalism of $f(R,T,Q)$ gravity while in section \textbf{3}, we
derive the equations for the stellar structure and construct the
system of differential equations using two EoS. Section \textbf{4}
explores physical attributes of considered compact stars while the
stability of our stellar system is analyzed in section \textbf{5}.
In the final section, we discuss the obtained results.

\section{Basic Formulation of $f(R,T,Q)$ Gravity}

The construction of $f(R,T,Q)$ gravity is developed on the strong
basis of non-minimal coupling of matter content and geometry. The
modified action of this gravity along with matter Lagrangian
$\mathcal{L}_{m}$ is given as \cite{11}
\begin{equation}\label{1}
A=\frac{1}{2\kappa^{2}}\int\sqrt{-g}\left({ f(R,T,Q)+
\mathcal{L}_{m}}\right)d^4x,
\end{equation}
where $\kappa^{2}(=1)$ is coupling constant and $g$ is the
determinant of the metric tensor $(g_{ab})$. The standard
energy-momentum tensor whose matter action based only on $g_{ab}$ is
defined by \cite{29}
\begin{equation}\label{2}
T_{ab}=-\frac{2}{\sqrt{-g}}\frac{\delta(\sqrt{-g}
\mathcal{L}_m)}{\delta g^{ab}}=g_{ab}
\mathcal{L}_{m}-\frac{2\partial\mathcal{L}_{m}}{\partial g^{ab}}.
\end{equation}
The field equations are obtained by applying the variation on action
(\ref{1}) w.r.t $g_{ab}$ as
\begin{eqnarray}\nonumber
&&R_{ab}f_{R}-\left(\frac{1}{2}f-\mathcal{L}_{m} f_{T}-\frac{1}{2}
\nabla_{\mu}\nabla_{\nu}(f_{Q}T^{\mu\nu})\right)g_{ab}+\left(g_{ab}
\Box-\nabla_{a}\nabla_{b}\right)f_{R}\\\nonumber&+&2f_{Q}
R_{\mu(a}T_{b)}^{\mu}-\nabla_{\mu}\nabla_{(a}[T_{b)}^{\mu}f_{Q}]
+\frac{1}{2}\Box(f_{Q}T_{ab})+\frac{1}{2}Rg_{ab}\mathcal{L}_{m}
f_{Q}-R_{ab}\mathcal{L}_{m}f_{Q}\\\label{3}&-&2(f_{T}g^{\mu\nu}
+f_{Q}R^{\mu\nu})\frac{\partial^{2}\mathcal{L}_{m}} {\partial
g^{ab}\partial g^{\mu\nu}}=(1+f_{T}+\frac{1}{2}Rf_{Q})T_{ab},
\end{eqnarray}
where $f_{R}\equiv\frac{\partial f}{\partial R}$,
$f_{T}\equiv\frac{\partial f}{\partial T}$ and
$\Box\equiv\nabla^{a}\nabla_{a}$. It is interesting to mention here
that for vanishing $Q$, the field equations of $f(R,T)$ theory can
be regained which can be compressed further to obtain $f(R)$ field
equations for vacuum case. The covariant divergence of field
equations (\ref{3}) leads to
\begin{eqnarray}\nonumber
\nabla^{a}T_{ab}&=&\frac{2}{2(1+f_{T})+Rf_{Q}}
\Big[\nabla_{a}\left(f_{Q}R^{a\mu}T_{\mu b}\right)
+\nabla_{b}(\mathcal{L}_{m}f_{T})-G_{ab}\nabla^{a}
(f_{Q}\mathcal{L}_{m})\Big.\\\label{4}&-&
\Big.\frac{1}{2}(f_{Q}R_{\mu\nu}+f_{T}g_{\mu\nu}) \nabla_{b}
T^{\mu\nu}-\frac{1}{2}[\nabla^{a}(Rf_{Q})
+2\nabla^{a}f_{T}]T_{ab}\Big].
\end{eqnarray}
It is quoted here that conservation equation does not hold in this
modified theory similar to other theories of gravity having
matter-curvature coupling \cite{10}. The standard stress energy
tensor for isotropic fluid distribution is
\begin{equation}\label{5}
T_{ab}=(p+\rho)V_{a}V_{b}- pg_{ab},
\end{equation}
where $p$ and $\rho$ represent the pressure and energy density of
the fluid, respectively and $V_{a}=\sqrt{g_{00}}(1,0,0,0)$ denotes
four velocity in comoving coordinates which satisfies the condition
$V_{a}V^{a}=1$. For a perfect fluid, we have $\mathcal{L}_{m}=-p$
that leads to $\frac{\partial^{2}\mathcal{L}_{m}} {\partial
g^{ab}\partial g^{\mu\nu}}=0$ \cite{11}.

Stellar evolution is a fascinating phenomenon because it is
responsible for the construction of relativistic stellar objects. It
is well-known that GR faces difficulties in explaining the universe
at large scales beyond the solar system as well as at large
energies. It seems natural to modify the Einstein gravity towards
better understanding of gravity at a very large and very small
scales. Since, the small modifications of GR may constitute an
unstable system in the interior solution, a phenomenon referred to
as matter instability. It is found that this instability can be
removed by a convenient choice of the functional form of modified
theories \cite{12}. For a spherical stellar system, the modified
field equations have to be solved both in the external as well as
internal geometries of the celestial object. In the exterior, where
$T = 0$, the field equations of $f(R,T,Q)$ gravity reduce to the
$f(R)$ theory and their solutions behave as in GR. Nevertheless,
interior solutions of the spherical system in $f(R,T,Q)$ gravity may
lead to large instabilities. This issue can be resolved if we
consider a viable model of this theory which satisfies the following
conditions
\begin{equation}\nonumber
3f_{RR}+\left(\frac{T}{2}-T^{00}\right)f_{QR}\geq0, \quad
\frac{1+f_{T}+\frac{1}{2}Rf_{Q}}{f_{R}-f_{Q}\mathcal{L}_{m}}>0.
\end{equation}

The classification of functional forms regarding different
configurations of matter in $f(R,T,Q)$ gravity are presented as
\begin{eqnarray}\nonumber
f(R,T,Q)=\left\{\begin{array}{lll}
R+f(Q),\\f_{1}(R)+f_{2}(Q),\\f_{1}(R)+f_{1}(R)f_{2}(Q)
,\\f_{1}(R)+f_{2}(T)+f_{3}(Q).\quad
\end{array}\right.
\end{eqnarray}
In this paper, we choose the first class, i.e., $f(R,T,Q)=R+f(Q)$
for which we consider $f(Q)=\delta Q$ to study stable/unstable
configurations of compact objects. This is the simplest functional
form which describes the relation between geometry and matter only
through the strong coupling between the Ricci and stress tensors.
This model is first suggested by Haghani et al. \cite{11} and has
widely been used to study different cosmological issues. Haghani et
al. \cite{11} examined the evolution as well as dynamics of the
universe corresponding to this model and found that for $\delta>0$,
this model well describes the expanding and collapsing phases of
cosmos. The energy conditions as well as laws of thermodynamics are
analyzed in this model. Baffou et al. \cite{29a} observed the
stability of power law and de Sitter solutions for this model and
found that the higher derivatives present in this theory can explore
a new aspect regarding early phases of cosmic evolution. The
physical features of compact stars using Krori-Barua solution have
also been studied for this model \cite{27}. Substituting this model
along with $\mathcal{L}_{m}=-p$ in Eq.(\ref{3}), it follows that
\begin{eqnarray}\nonumber
G_{ab}&=&\frac{1}{1+\delta p}\left[T_{ab}+\frac{\delta}{2}RT_{ab}
+\frac{\delta}{2}\left\{Qg_{ab}-\Box T_{ab}-
\nabla_{\mu}\nabla_{\nu}(T^{\mu\nu})g_{ab}\right\}\right.\\\label{6}&-&\left.2\delta
R_{\mu(a}T_{b)}^{\mu}+\delta\nabla_{\mu}\nabla_{(a}T_{b)}^{\mu}\right],
\end{eqnarray}
where $G_{ab}$ is the usual Einstein tensor. Also, for the
considered model, the non-conservation equation of the stress-energy
tensor reduces to
\begin{eqnarray}\nonumber
\nabla^{a}T_{ab}&=&\frac{2\delta}{2+\delta R}
\Big[\nabla_{a}\left(R^{a\mu}T_{\mu b}\right)+G_{ab}\nabla^{a}(
p)-\frac{1}{2}(R_{\mu\nu})\nabla_{b}
T^{\mu\nu}-\frac{1}{2}\nabla^{a}(R)T_{ab}\Big].\\\label{7}
\end{eqnarray}

\section{Equations of Stellar Structure}

It is assumed that a star remains in a steady state described by
static spherically symmetric metric given by
\begin{equation}\label{8}
ds^{2}_{-}=e^{\mu(r)}dt^{2}-e^{\lambda(r)}dr^{2}-r^{2}(d\theta^{2}+\sin^{2}\theta
d\phi^{2}).
\end{equation}
For this interior geometry, the field equations (\ref{6}) along with
matter content (\ref{5}) turn out to be
\begin{eqnarray}\nonumber
e^{-\lambda}\left(\frac{\lambda'}{r}
-\frac{1}{r^{2}}\right)+\frac{1}{r^{2}}&=&\frac{1}{1+\delta
p}\left[\rho+\frac{\delta}{2}\left\{\rho''-p''-\frac{2e^{\lambda}}{r^{2}}
\left(\rho-p\right)+(\frac{4}{r}-\frac{\lambda'}{2})
\right.\right.\\\nonumber&\times&\left.\left.(\rho'-p')
+\rho\left(\mu''+2\mu'^{2}-\mu'\lambda'+\frac{2}{r^{2}}
-\frac{2\lambda'}{r}\right)+p\right.\right.\\\label{9}&\times&
\left.\left.\left(\frac{\mu''}{2}-\frac{\mu'}{r}
-\frac{2}{r^{2}}-\frac{3\mu'\lambda'}{4}+\frac{2\lambda'}{r}
+\frac{7\mu'^{2}}{4}\right)e^{-\lambda}\right\}\right],\\\nonumber
e^{-\lambda}\left(\frac{\mu'}{r}+\frac{1}{r^{2}}\right)
-\frac{1}{r^{2}}&=&\frac{1}{1+\delta p}\left[p\left(1
-\frac{\delta}{r^{2}}\right) +\frac{\delta e^{-\lambda}}{2}
\left\{\rho\left(\frac{\mu''}{2}+\frac{3\mu'^{2}}{4}
+\frac{\mu'}{r}\right.\right.\right.\\\nonumber&+&\left.
\left.\left.\frac{\mu'\lambda'}{4}\right)
+p\left(\frac{\mu'^{2}}{4}-\frac{\mu''}{2}+\frac{3\mu'}{r}
-\frac{2}{r^{2}}+\frac{5\mu'\lambda'}{4}+\frac{2\lambda'}{r}
\right.\right.\right.\\\label{10}&+&\left.\left.\left.\lambda'^{2}
+\lambda''\right)+p'\left(\frac{\mu'}{2}+\frac{6}{r}
+2\lambda'\right)\right\}\right],\\\nonumber
\end{eqnarray}
where prime reveals derivative w.r.t radial coordinate. Moreover,
from the non-conservation of the stress energy tensor (\ref{7}), the
hydrostatic equilibrium equation is calculated as
\begin{eqnarray}\nonumber
p'+(\rho+p)\frac{\mu'}{2}&=&\frac{\delta}{2+\delta
R}\left[2p\left\{\left(\frac{\mu'}{2}+\frac{2}{r}\right)
\left(\frac{\lambda'}{r}+\frac{\mu'\lambda'}{4}-\frac{\mu'^{2}}{4}
-\frac{\mu''}{2}\right)+\frac{\lambda''}{r}\right.\right.\\\nonumber&-&
\left.\left.\frac{\mu'''}{2}-\frac{\mu'\mu''}{2}
+\frac{\mu'\lambda''}{4}+\frac{\mu''\lambda'}{2}
+\frac{\mu'^{2}\lambda'}{8}-\frac{\lambda'}{r^{2}}+\frac{\mu'}{r^{2}}
-\frac{\mu'\lambda'^{2}}{8}\right.\right.\\\nonumber&-&
\left.\left.\frac{\lambda'^{2}}{2r}+\frac{2}{r^{3}}\right\}
+pe^{\lambda}\left\{\frac{3\mu''\lambda'}{2}-\mu'''
-\frac{4e^{\lambda}}{r^{3}}-\mu'\mu''+\frac{\mu'^{2}\lambda'}{2}
\right.\right.\\\nonumber&+&\left.\left.\frac{\mu'\lambda''}{2}
-\frac{\mu'\lambda'^{2}}{2}+\frac{4\lambda'}{r^{2}}
-\frac{2\lambda''}{r}+\frac{2\lambda'^{2}}{r}+\frac{2\mu'^{2}}{r}
-\frac{2\mu''}{r}+\frac{2\mu'\lambda'}{r}\right\}
\right.\\\nonumber&+&\left.\frac{2p'}{r^{2}}\left\{e^{\lambda}
(\mu'r-e^{\lambda})+1-\frac{r\lambda'}{2}+\frac{\mu'r}{2}\right\}
-\left(\rho'+\mu'\rho\right)\left(\frac{\mu''}{2}
\right.\right.\\\label{11}&+&\left.\left.\frac{\mu'}{r}
-\frac{\mu'\lambda'}{4}+\frac{\mu'^{2}}{4}\right)\right].
\end{eqnarray}
It is interesting to note that field equations (\ref{9}), (\ref{10})
and conservation equation (\ref{11}) reduce to standard GR equations
\cite{30,31} for $\delta=0$. From Eqs.(\ref{9})-(\ref{11}), we
obtain a set of three differential equations as
\begin{eqnarray}\nonumber
p''&=&p'\left(\frac{\lambda'}{2}-\frac{4}{r}\right)
+p\left(\frac{\mu''}{2}-\frac{\mu'}{r}+\frac{7\mu'^{2}}{4}
-\frac{3\mu'\lambda'}{4}+\frac{2e^{\lambda}}{r^{2}}\left(\frac{1
-\delta}{\delta}\right)\right)+\rho''\\\nonumber&+&\rho'\left(\frac{4}{r}
-\frac{\lambda'}{2}\right)+\rho\left(2e^{\lambda}\left(\frac{1}{\delta}
-\frac{1}{r^{2}}\right)+\mu''+2\mu'^{2}-\mu'\lambda'
+\frac{2}{r^{2}}-\frac{2\lambda'}{r}\right),\\\label{12}\\\nonumber
\lambda''&=&-\lambda'\left(\lambda'+\frac{2}{r}+\frac{2p'}{p}
+\frac{4\mu'}{4}+\frac{\mu'\rho}{4p}\right)
-\frac{2e^{\lambda}}{\delta}\left(\frac{1}{pr^{2}}
+\frac{1}{\delta}\right)+2\left(\frac{1+\delta p}{\delta
p}\right)\\\nonumber&\times&\left(\frac{\mu'}{r}
+\frac{1}{r^{2}}\right)-\frac{\rho}{p}\left(\frac{\mu''}{2}
+\frac{\mu'}{r} +\frac{3\mu'^{2}}{4}\right)+\frac{\mu''}{2}
-\frac{2}{r^{2}}-\frac{\mu'^{2}}{4}-\frac{3\mu'}{r}
\\\label{13}&-&\frac{p'}{p}\left(\frac{\mu'}{2}
+\frac{6}{r}\right),\\\nonumber \mu'''&=&\frac{1}{1
+e^{\lambda}}\left[\mu''\left\{-e^{-\lambda}
\left(\frac{p'}{p}+\frac{\mu'(\rho+p)}{2p}\right)-\frac{3\mu'}{2}
-\frac{2}{r}+\lambda'+e^{\lambda}\left(\frac{3\lambda'}{2}
-\mu'\right.\right.\right.\\\nonumber&-&\left.\left.\left.
\frac{2}{r}\right)-\frac{\rho'+\mu'\rho}{2\delta
p}\right\}+\mu'\left\{e^{-\lambda}\left(\frac{p'}{p}
\left(\frac{\lambda'}{2}-\frac{2}{r}-\frac{\mu'}{2}\right)
+\frac{\mu'\lambda'}{4p}+\left(\rho+p\right)\right.\right.\right.
\\\nonumber&\times&\left.\left.\left.\left(\frac{\lambda'}{pr}
-\frac{\mu'}{pr}-\frac{1}{pr^{2}}-\frac{\mu'^{2}}{4p}\right)\right)
-\frac{\rho+p}{pr^{2}}+\frac{\mu'\lambda'}{2}
+\frac{2\lambda'}{r}-\frac{\mu'^{2}}{4}-\frac{\mu'}{r}
+\frac{\lambda''}{2}\right.\right.\\\nonumber&+&\left.\left.
\frac{2}{r^{2}}-\frac{\lambda'^{2}}{4}
+\frac{p'}{pr}+e^{\lambda}\left(\frac{\lambda''}{2}
+\frac{\mu'\lambda'}{2}-\frac{\lambda'^{2}}{2}
+\frac{2}{r^{2}}+\frac{2\lambda'}{r}+\frac{2p'}{pr^{2}}\right)\right\}
\right.\\\nonumber&-&\left.\left(\frac{\rho'+\mu'\rho}{\delta
p}\right)\left(\frac{\mu'}{r}-\frac{\mu'\lambda'}{4}
+\frac{\mu'^{2}}{4}\right)+\frac{2p'}{pr^{2}}\left(2
-e^{2\lambda}-2e^{-\lambda}-\frac{r\lambda'}{2}\right)
\right.\\\nonumber&+&\left.\frac{2p'}{p}\left
(\frac{\lambda'e^{-\lambda}}{r}-\frac{1}{\delta}\right)
+e^{\lambda}\left(\frac{4\lambda'}{r^{2}}
-\frac{2\lambda''}{r}+\frac{2\lambda'^{2}}{r}
-\frac{4e^{\lambda}}{r^{3}}\right)\right.
\\\label{14}&+&\left.\frac{2\lambda'}{r^{2}}
+\frac{2\lambda''}{r}+\frac{4}{r^{3}}
-\frac{\lambda'^{2}}{r}\right].
\end{eqnarray}
Since we have a set consisting of three non-linear differential
equations in four unknowns $p,~\rho,~\lambda$ and $\mu$, so we need
EoS which will be helpful to reduce one unknown parameter.

To solve the system of differential equations, we suppose a direct
and systematic relation between pressure and density of the fluid
which depicts the form of matter for the proposed set of physical
constraints named as EoS. In compact stars, white dwarfs have masses
about $1.4M_{\odot}$ and their radii are much smaller than the sun.
On the other hand, the neutron stars can possess larger masses up to
$3M_{\odot}$ \cite{32}. The attractive gravitational effect in white
dwarfs is dominated by the degeneracy pressure of electrons whereas
in neutron stars, this balance is maintained by the degeneracy
pressure of neutrons. Neutron stars are the utmost curious objects
and can be further transformed into the black hole if they possess
highly dense core whereas the slighter dense cores in neutron stars
collapse into a quark star. The transition of quark stars from
neutron stars has been discussed in literature \cite{33}. The
imaginary forms of compact objects formulated by three flavors (up,
down and strange) are referred to as quark stars. The structure of
these stars are smaller in size as well as highly dense and maintain
an extreme gravitational field. In neutron stars as well as white
dwarfs, pressure against the gravitational pull has the same source
known as quantum pressure.

To define the polytropic stars, the polytropic EoS is given by
\begin{equation}\nonumber
p=\alpha\rho^{1+\frac{1}{n}},
\end{equation}
where $n$ is a polytropic index and $\alpha$ is a polytropic
constant. In literature \cite{34}-\cite{35a}, it is found that the
realistic EoS can be obtained through piecewise polytropic EoS with
index $n\in[0.5,1]$. To examine physical features of stellar models
in $f(R,T,Q)$ gravity, we consider MIT bag model given by the
relation $p = 1/3(\rho-4\mathfrak{B})$, where $\mathfrak{B}$
symbolizes the bag constant (for quark stars) and realistic
polytropic EoS $p=\alpha\rho^{2}$ (corresponds to neutron stars).
These EoS have been used successfully for the analysis of stellar
configuration of compact stars \cite{26,28a,28b,34}-\cite{35a}.

\section{Physical Properties of Compact Stars}

In this section, we analyze the physical consistency of stellar
equations to observe the properties of compact stars. With the help
of EoS, we solve the stellar structure equations numerically using
initial conditions for different values of $\delta$. In the interior
of stellar models, the density and pressure should be positive,
finite as well as regular at all points. Applying this behavior at
the center $r = 0$, we obtain the following initial conditions from
the field equations
\begin{eqnarray}\nonumber
&&\mu''(0)=0,\quad \mu'(0)=0,\quad \mu(0)=0, \quad
\lambda'(0)=0,\\\label{15} &&\lambda(0)=0,\quad p(0)=p_{c}, \quad
\rho(0)=\rho_{c},\quad p'(0)=0,
\end{eqnarray}
where $\rho_{c}$ and $p_{c}$ represent some central values which we
fix for numerical analysis. Also, consideration of EoS decreases one
condition such that we need only the value of $p(0)$. In this paper,
we take $\mathfrak{B}=60 MeV/fm^{3}$ \cite{26}, $\alpha=4.78028 ×
10^{-5} [fm^{3}/MeV]^{2/3}$ and $p_{c}=200 MeV/fm^{3}$ \cite{28a}.
Throughout this analysis, we are employing the units of mass as
$M_{\odot}$, density (pressure) as $MeV/fm^{3}$ and radius as $km$
\cite{26,28a}.

\subsection{Metric Functions, Density and Pressure}

Here, we examine the role of metric functions, energy density and
pressure for both quark as well as polytropic compact stars for
different values of $\delta$. The variation of physical parameters
$e^{\mu}$, $e^{\lambda}$, $p$ and $\rho$ versus radial coordinate
are featured in Figures \textbf{1}-\textbf{4}. From Figure
\textbf{1}, it is observed that with the increasing value of
$\delta$, the metric function $e^{\mu}$ decreases for both MIT bag
model and polytropic EoS. The graphical behavior of other metric
function $e^{\lambda}$ also reduces for both EoS with the increment
in the model parameter $\delta$ as shown in Figure \textbf{2}.
However, the behavior of these metric potentials is positive
definite which shows that our set of stellar system vanishes any
kind of singularity.
\begin{figure}\center
\epsfig{file=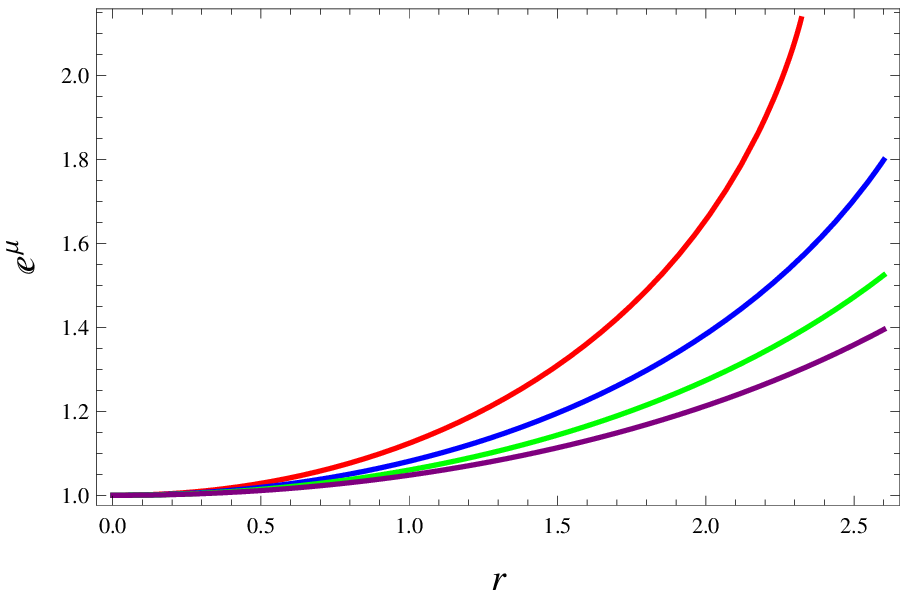,width=0.45\linewidth}
\epsfig{file=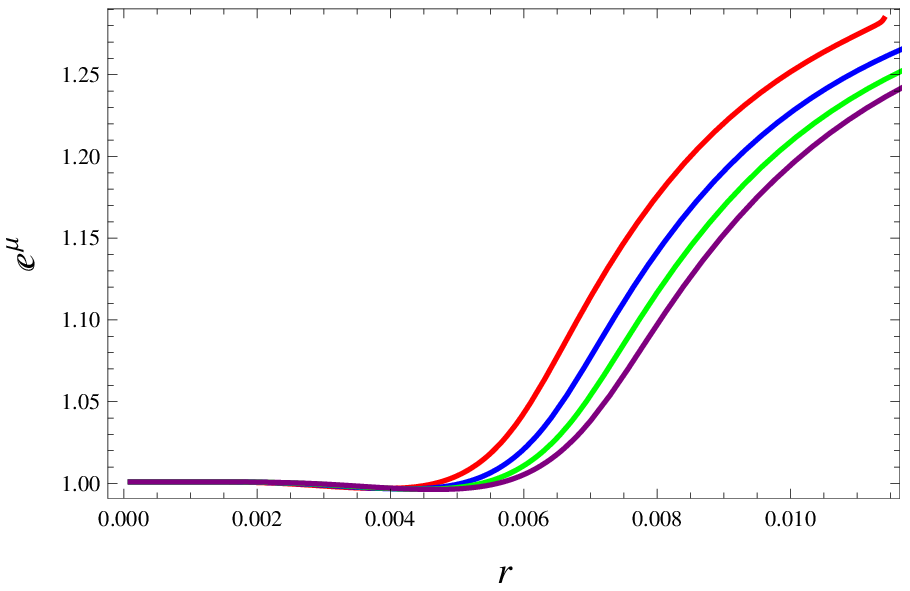,width=0.45\linewidth}\\
\caption{Variation of $e^{\mu}$ versus $r$ with MIT bag model (left)
and polytropic EoS (right) for $\delta=2$ (red), $\delta=3$ (blue),
$\delta=4$ (green) and $\delta=5$ (purple).}
\end{figure}
\begin{figure}\center
\epsfig{file=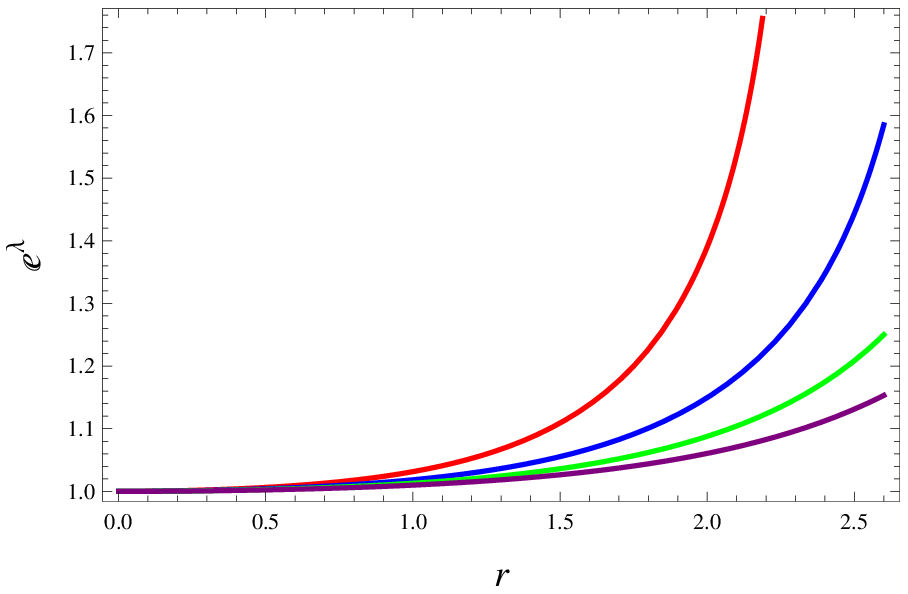,width=0.45\linewidth}
\epsfig{file=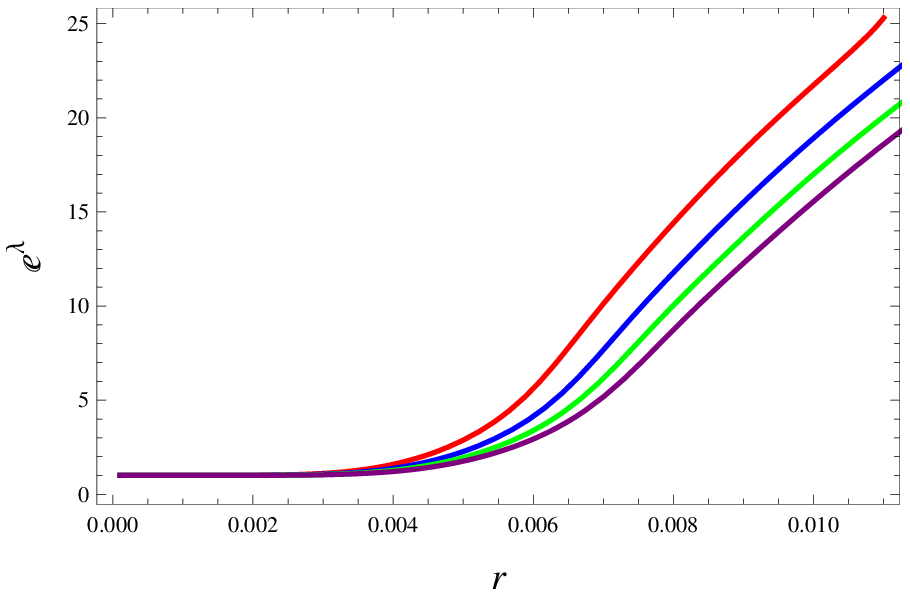,width=0.45\linewidth}\\
\caption{Variation of $e^{\lambda}$ versus $r$ with MIT bag model
(left) and polytropic EoS (right) for $\delta=2$ (red), $\delta=3$
(blue), $\delta=4$ (green) and $\delta=5$ (purple).}
\end{figure}

In the interior of dense compact stars, the pressure as well as
energy density should possess maximum value at the center. Figures
\textbf{3} and \textbf{4} exhibit that these physical quantities
have maximum behavior at $r=0$ and decrease towards the surface of
compact objects. It is also observed that the values of these matter
variables increase with increasing values of $\delta$ which depicts
the existence of dense cores of compact stars under the strong
influence of coupling term $Q$. From the graphical analysis, it is
found that the effect of pressure corresponding to quark stars is
zero for $\delta=2$ at $r\approx 2.33km$ whereas using realistic
polytropic EoS, it vanishes at $r\approx 0.014km$ for the same value
of coupling parameter. It is worth mentioning here that the radius
obtained for quark stars is physically viable while the radius
observed for realistic polytropic EoS is much smaller than the
predicted radii of neutron stars. However, in the context of
$f(R,T)$ gravity, the physical attributes of the stellar system with
radius $r\approx 0.06km$ have been discussed in literature for
realistic polytropic EoS \cite{35b}.\begin{figure}\center
\epsfig{file=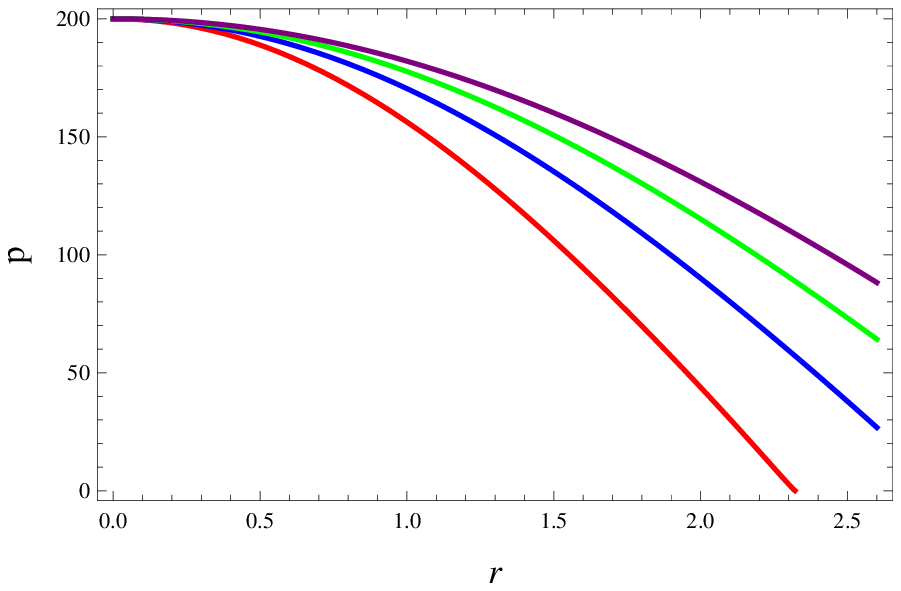,width=0.45\linewidth}
\epsfig{file=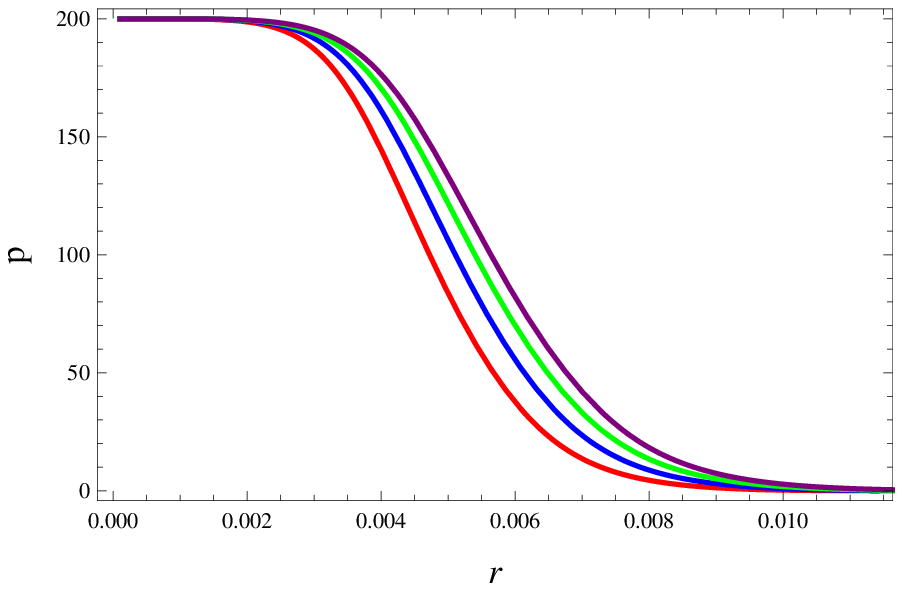,width=0.45\linewidth}\\
\caption{Plot of $p$ versus $r$ with MIT bag model (left) and
polytropic EoS (right) for $\delta=2$ (red), $\delta=3$ (blue),
$\delta=4$ (green) and $\delta=5$ (purple).}
\end{figure}
\begin{figure}\center
\epsfig{file=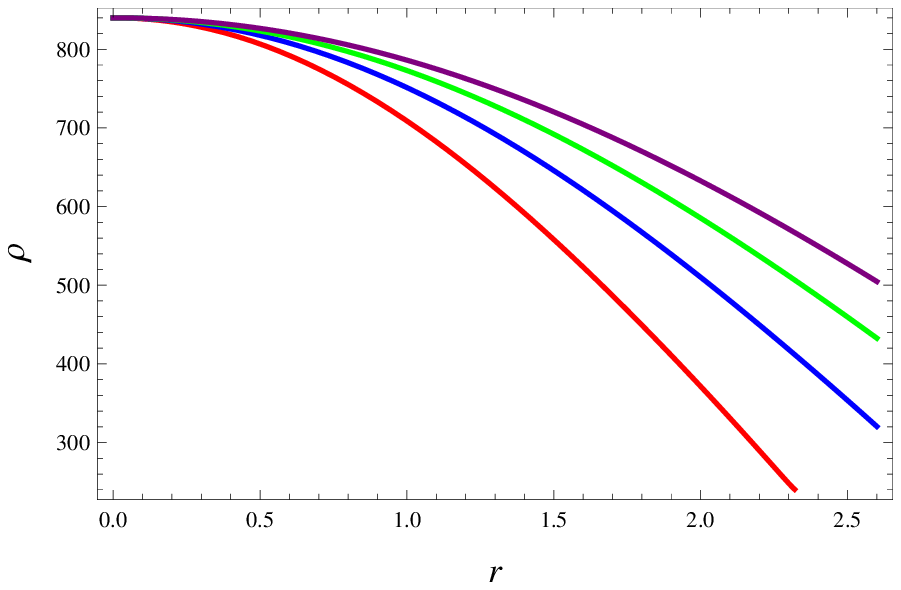,width=0.45\linewidth}
\epsfig{file=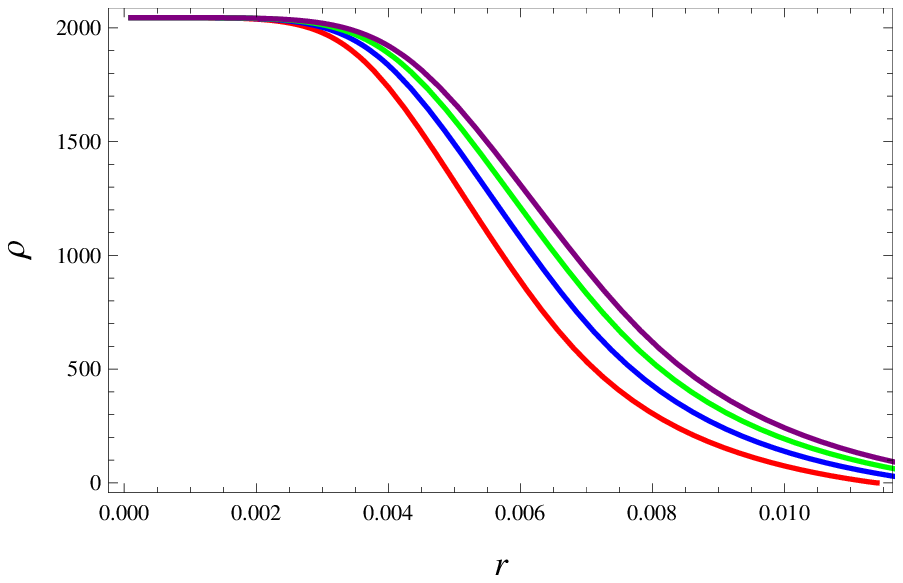,width=0.45\linewidth}\\
\caption{Plot of $\rho$ versus $r$ with MIT bag model (left) and
polytropic EoS (right) for $\delta=2$ (red), $\delta=3$ (blue),
$\delta=4$ (green) and $\delta=5$ (purple).}
\end{figure}

\subsection{Energy Conditions}

To examine the nature of normal or exotic matter in the interior
geometry of compact objects, energy conditions play a vital role.
The consistency of these conditions are governed by satisfying the
following inequalities \cite{36}
\begin{itemize}
\item NEC:  \quad $\rho+ p\geq 0$,
\item SEC:  \quad $\rho+ p\geq 0,\quad \rho+ 3p\geq0$,
\item DEC:  \quad $\rho\geq 0,\quad \rho-p\geq 0$,
\item WEC:  \quad $\rho+ p\geq 0,\quad \rho\geq 0$.
\end{itemize}
Figure \textbf{5} indicates that our system of differential
equations, adopting two EoS, is feasible with all energy conditions
for different values of $\delta$ which also affirms the physical
acceptability of particular functional form of this
gravity.\begin{figure}\center
\epsfig{file=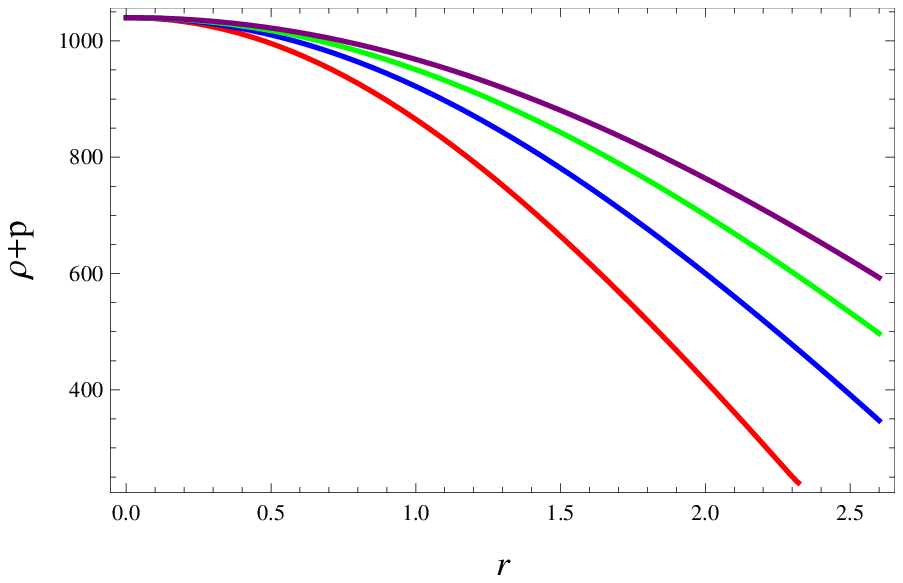,width=0.45\linewidth}
\epsfig{file=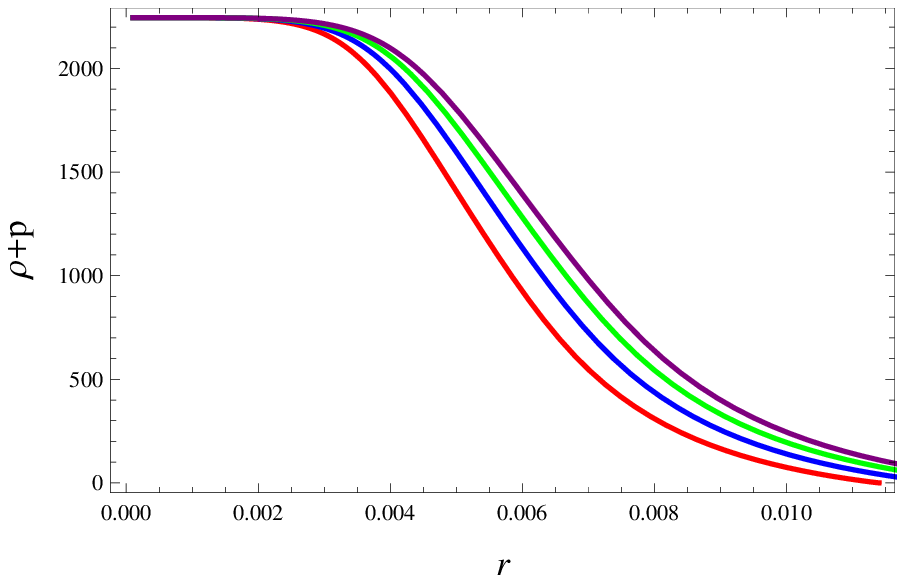,width=0.45\linewidth}
\epsfig{file=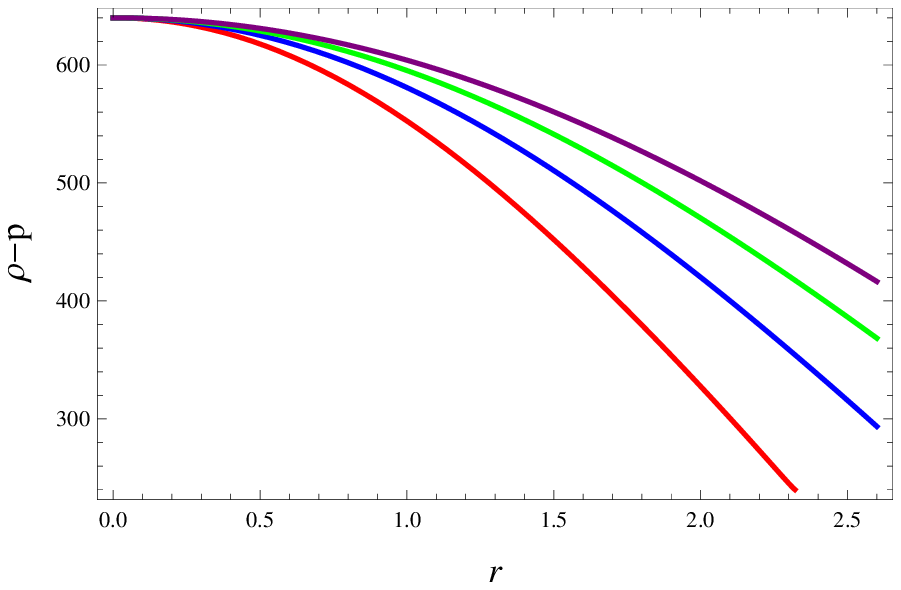,width=0.45\linewidth}
\epsfig{file=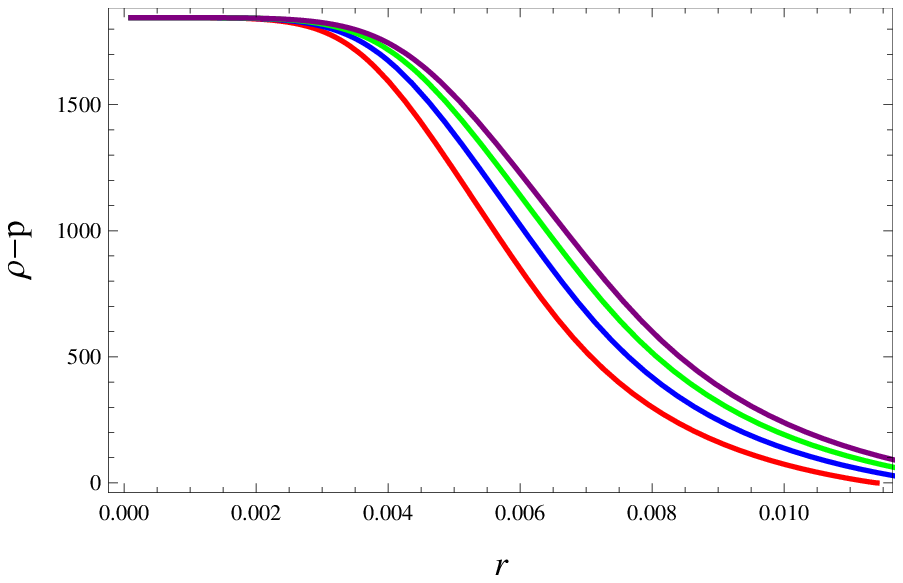,width=0.45\linewidth}
\epsfig{file=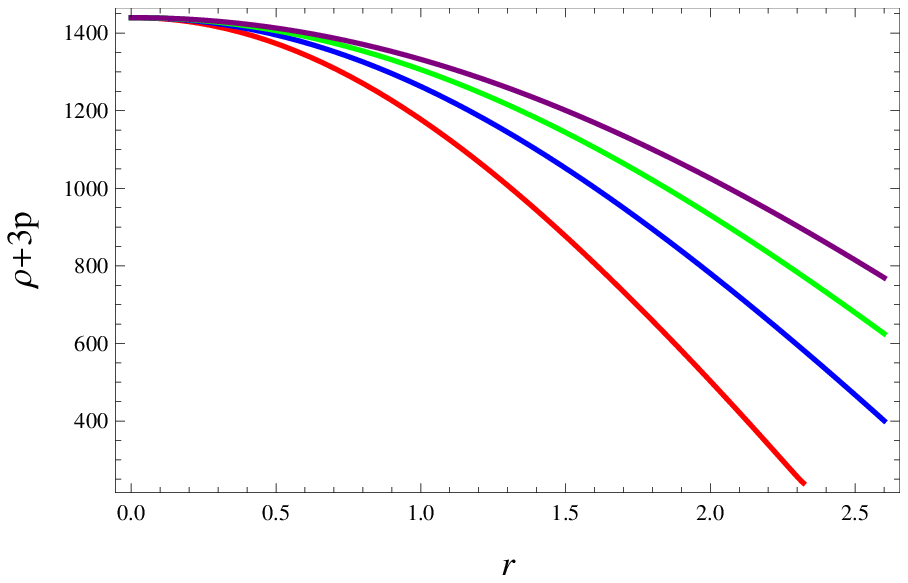,width=0.45\linewidth}
\epsfig{file=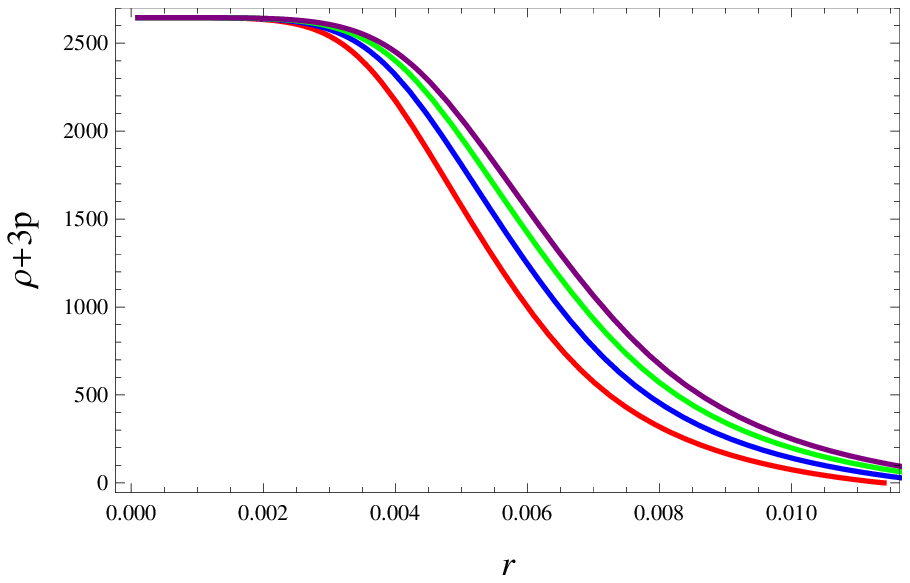,width=0.45\linewidth}\\
\caption{Behavior of energy conditions with MIT bag model (left) and
polytropic EoS (right) for $\delta=2$ (red), $\delta=3$ (blue),
$\delta=4$ (green) and $\delta=5$ (purple).}
\end{figure}
\subsection{Mass-Radius Relation and Surface Redshift}

Here, we investigate mass as well as surface redshift of the compact
stars for considered EoS with radial coordinate. At the boundary of
the star's surface, i.e., at $r=R$, the interior solution of stellar
objects connects smoothly with the Schwarzschild vacuum solution.
The Schwarzschild approach has widely been used from different
possibilities of the matching conditions to explore the features of
stellar compact objects \cite{37}. At the boundary ($r=R$), the
interior as well as exterior geometries are linked together by the
following relation
\begin{equation}\label{16}
e^{\lambda(R)}=\left(1-\frac{2M}{R}\right)^{-1},
\end{equation}
where $M$ reveals the total mass of stellar objects. The expression
of $M$ can be calculated from Eq.(\ref{16}) as
\begin{equation}\nonumber
M=\frac{R}{2}\left(1-e^{-\lambda(R)}\right).
\end{equation}
The surface redshift is also a significant phenomenon to interpret
the dynamics of strong interaction between internal distribution of
a star and its EoS. The formula for $z_{s}$ corresponding to
mass-function is defined as
\begin{equation}\nonumber
z_{s}=\left(1-\frac{2M}{R}\right)^{-1/2}-1.
\end{equation}
\begin{figure}\center
\epsfig{file=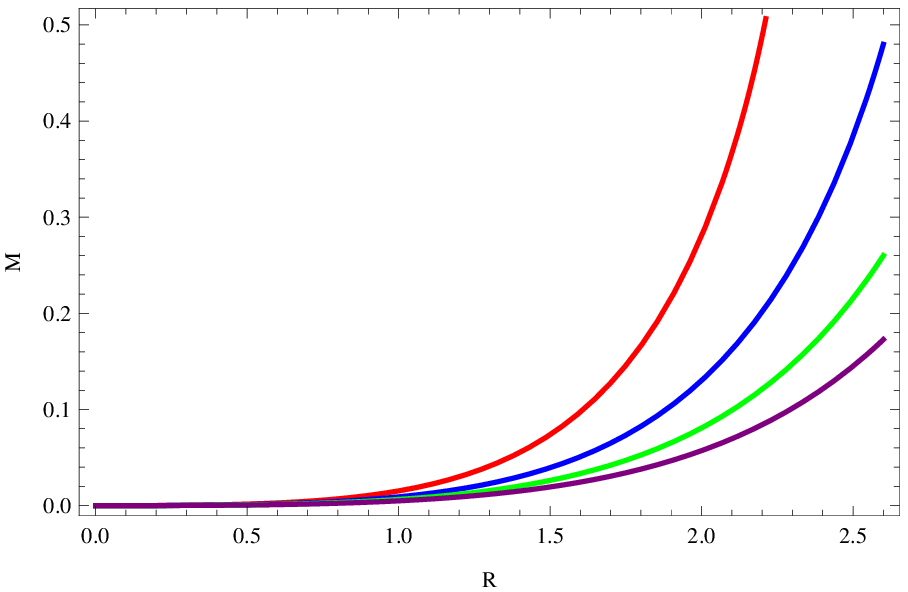,width=0.45\linewidth}
\epsfig{file=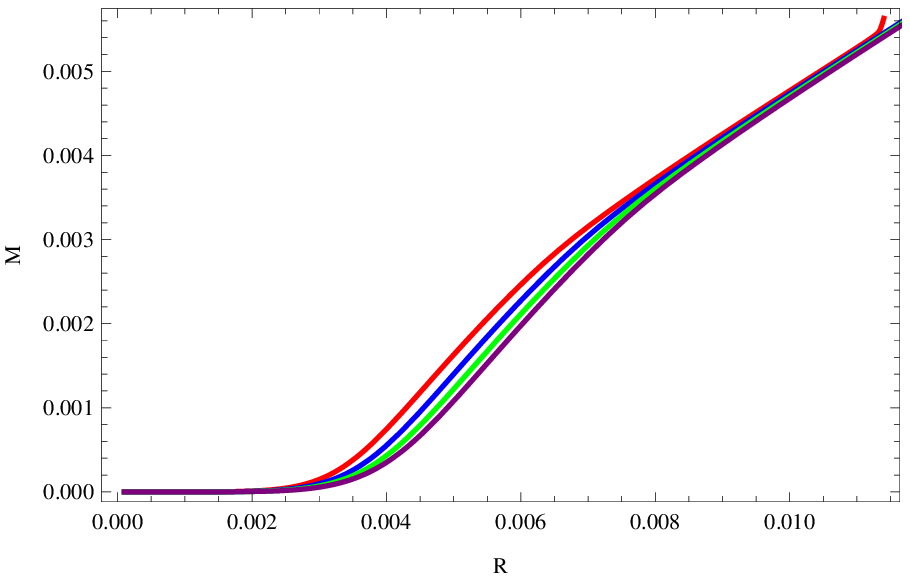,width=0.45\linewidth}\\
\caption{Mass-radius relation with MIT bag model (left) and
Polytropic EoS (right) for $\delta=2$ (red), $\delta=3$ (blue),
$\delta=4$ (green) and $\delta=5$ (purple).}
\end{figure}

For static spherically symmetric configurations with perfect fluid
distribution, the maximum bound for the surface redshift parameter
is obtained as $z_{s}\leq 2$ \cite{6}. The graphical behavior of
mass-radius relation as well as redshift parameter is presented in
Figures \textbf{6} and \textbf{7}. Figure \textbf{6} shows that the
value of mass-function decreases with increasing values of $\delta$.
In $f(R,T)$ scenario, it is observed that the maximum mass
$2.17M_{\odot}$ and $z_{s}\approx0.4$ are obtained for MIT bag model
\cite{28}. In our case for $\delta=2$, the maximum mass
$0.5M_{\odot}$ and $0.006M_{\odot}$ are attained for quark and
polytropic stars, respectively. It is also found that the bound of
surface redshift is in good agreement with the required bound for
both EoS as shown in Figure \textbf{7}.
\begin{figure}\center
\epsfig{file=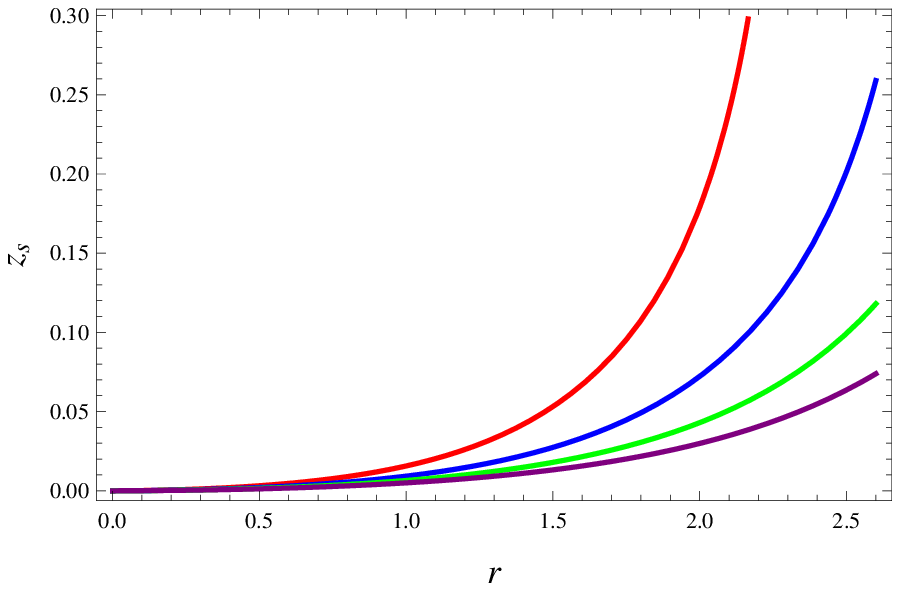,width=0.45\linewidth}
\epsfig{file=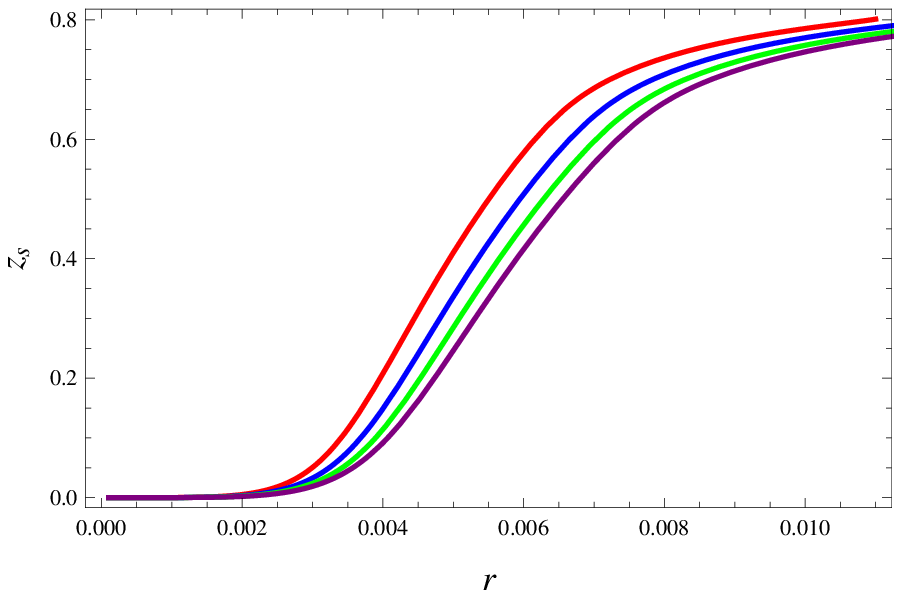,width=0.45\linewidth}\\
\caption{Plot of redshift versus $r$ with MIT bag model (left) and
Polytropic EoS (right) for $\delta=2$ (red), $\delta=3$ (blue),
$\delta=4$ (green) and $\delta=5$ (purple).}
\end{figure}

\section{Stability of Compact Objects}

The stability of relativistic structure has great importance in
analyzing physically acceptable models. We investigate the stability
of compact stars using two EoS by examining causality condition and
adiabatic index.

\subsection{Causality Condition}

Here, we are interested to check the speed of sound $(v_{s}^{2})$
using Herrera's cracking concept \cite{38}. The causality condition
demands that the squared sound speed represented by
$v_{s}^{2}=dp/d\rho$ should be in the interval $[0,~1]$, i.e.,
$0\leq v_{s}^{2}\leq1$ everywhere inside the stellar model for a
physically acceptable object. Figure \textbf{8} reveals that for
quark stars, we have a constant required value of $v_{s}^{2}$ which
is completely matched with the observed value of speed of sound in
$f(R,T)$ gravity \cite{28} while for polytropic EoS, the value of
speed of sound lies within the required range and it is also
monotonically decreasing towards the surface of stars. Hence, our
system of stellar equations is consistent with the required
causality condition for both EoS and shows stable structure for all
considered values of the coupling parameter.
\begin{figure}\center
\epsfig{file=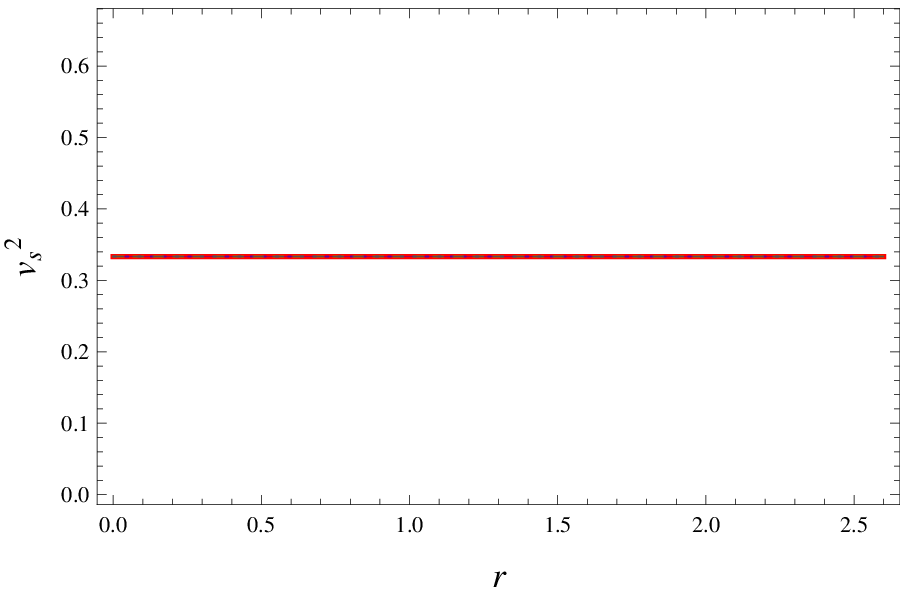,width=0.45\linewidth}
\epsfig{file=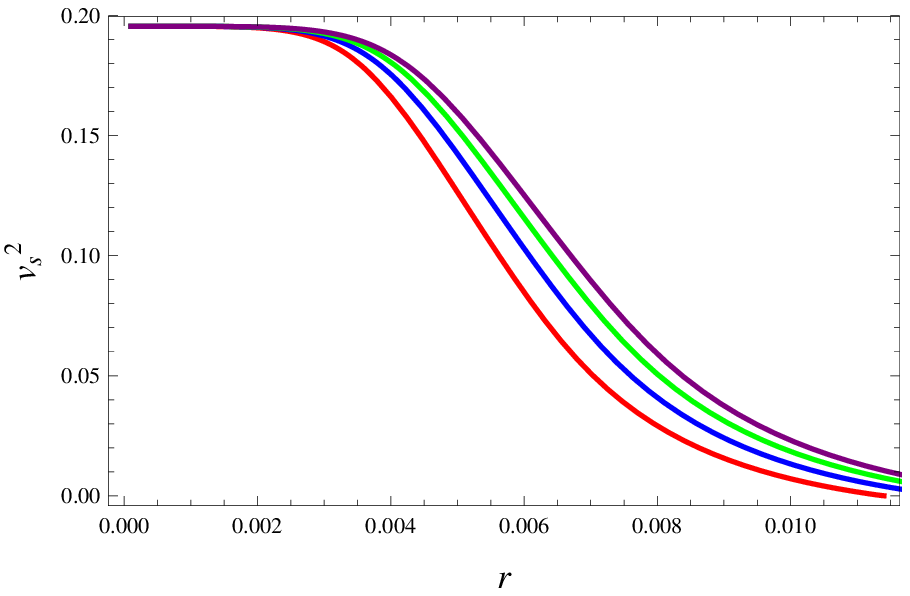,width=0.45\linewidth}\\
\caption{Behavior of speed of sound versus $r$ with MIT bag model
(left) and Polytropic EoS (right) for $\delta=2$ (red), $\delta=3$
(blue), $\delta=4$ (green) and $\delta=5$ (purple).}
\end{figure}

\subsection{Adiabatic Index ($\Gamma$)}

The stiffness of the EoS corresponding to considered energy density
is described by the adiabatic index which plays a major role to
analyze the stability of stellar models. Chandrasekhar (as a
pioneer) \cite{39} and many researchers \cite{40} inspected the
dynamical stability of the stellar system. It is predicted that the
adiabatic index must be $>\frac{4}{3}$ for a dynamically stable
stellar object \cite{40}. The expression for adiabatic index is of
the form
\begin{equation}\nonumber
\Gamma=\frac{\rho+p}{p}\left(dp/d\rho\right)=\frac{\rho+p}{p}
\left(v_{s}^{2}\right).
\end{equation}
The graphical interpretation of the adiabatic index is exhibited in
Figure \textbf{9} for both EoS. This depicts that the set of stellar
equations is dynamical stable for all values of the coupling
parameter $\delta$ as the values of $\Gamma>\frac{4}{3}$ for both
quark as well as realistic polytropic stars.\begin{figure}\center
\epsfig{file=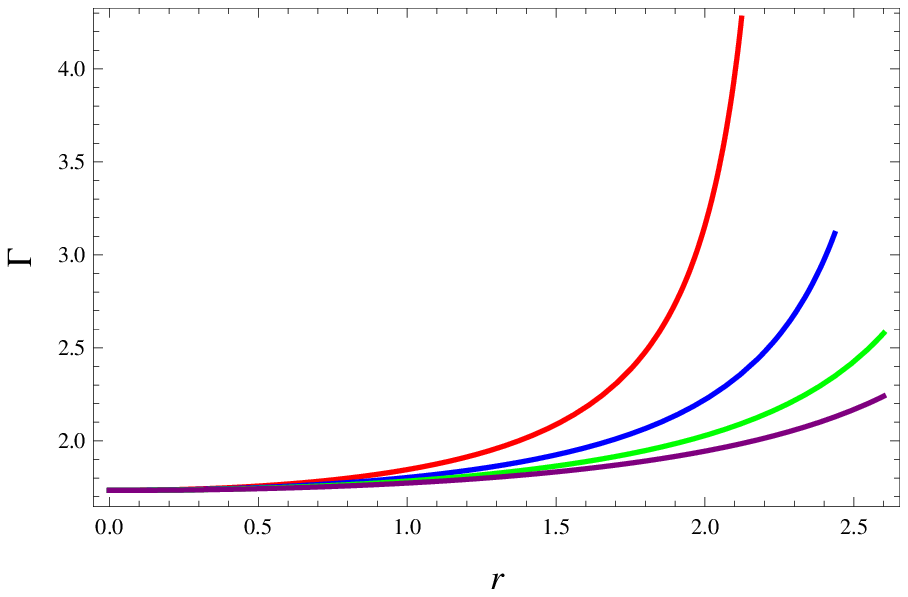,width=0.45\linewidth}
\epsfig{file=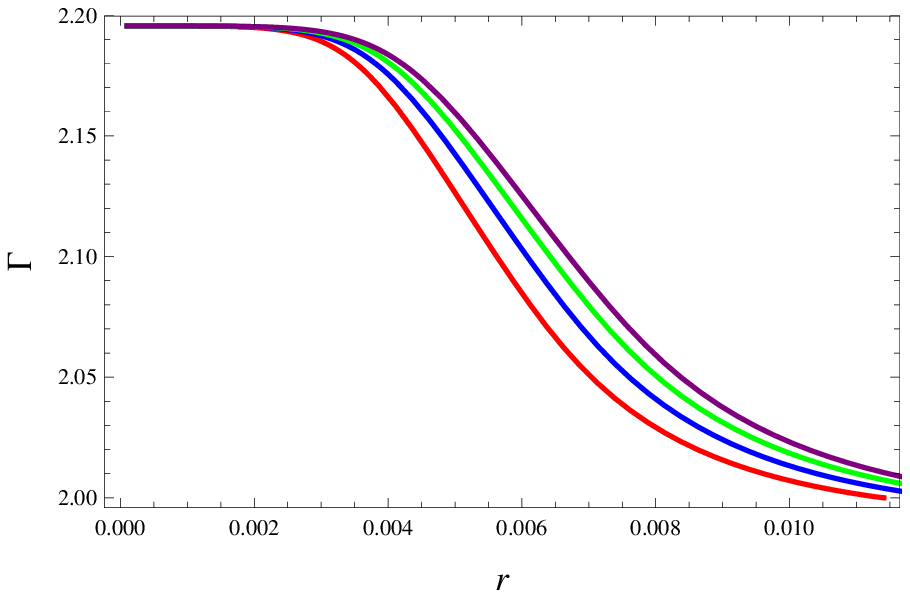,width=0.45\linewidth}\\
\caption{Plots of adiabatic index versus $r$ with MIT bag model
(left) and polytropic EoS (right) for $\delta=2$ (red), $\delta=3$
(blue), $\delta=4$ (green) and $\delta=5$ (purple).}
\end{figure}

\section{Discussion and Conclusions}

This paper explores the physical attributes as well as dynamical
stability of compact objects with two EoS (MIT bag model and
realistic polytropic) in the framework of $f(R,T,Q)$ gravity. The
physical properties of proposed compact stars have been observed by
deriving equations of stellar structure and hydrostatic equilibrium
equation for $R+\delta Q$ gravity model. The hydrostatic equilibrium
equation, also known as Tolman-Oppenheimer-Volkoff (TOV) equation,
is an extension due to the presence of extra terms coming from
$\delta Q$. The stellar configurations of quark and polytropic stars
have been analyzed for different values of the model parameter
$\delta$.

We have constructed a system of differential equations and derived
the initial conditions required for the numerical analysis. The
variation of metric functions (Figures \textbf{1} and \textbf{2})
reveals that our system of stellar equations is free from any type
of geometrical or physical singularity. The regularity conditions
for energy density as well as pressure are also satisfied. It is
found that at the center, stars exhibit maximum pressure and density
which decrease monotonically towards the boundary of the stellar
objects. The radii of approximately $2.33km$ and $0.014km$ are
observed for quark and polytropic stars, respectively. In the
background of gravity, it is investigated that the radius of
realistic polytropic star is smaller than the inspected radius of
neutron stars. Our stellar system is also found to be consistent
with all energy bounds for all suggested values of $\delta$ (Figure
\textbf{5}).

The mass-radius relation as well as surface gravitational redshift
indicate that the maximum mass point for quark stars is
$0.5M_{\odot}$ while for polytropic stars is $0.006M_{\odot}$ at
$\delta=2$. The maximum surface redshift are approximately equal to
$0.3$ and $0.8$ for MIT bag model and polytropic EoS, respectively.
The graphical analysis predicts that with the increment in the
values of $\delta$, the mass of stars reduces and density in the
interior of stars gradually increases which suggests the existence
of utmost dense relativistic objects. We have obtained that the
speed of sound for all values of $\delta$ shows constant behavior
inside the system for quark stars and lies between $[0,~1]$ for both
quark as well as polytropic stars which confirms that the stars for
both EoS are stable. The behavior of adiabatic index in Figure
\textbf{9} shows that $\Gamma>\frac{4}{3}$ for both types of compact
stars, which verifies the stability against an infinitesimal radial
adiabatic perturbation.

In GR, there are some upper bounds on the masses of white dwarfs and
neutron stars. Accroding to Chandrasekhar \cite{39}, the maximum
mass limit of a white dwarf is $1.4M_{\odot}$ and the TOV limit
\cite{40} provides an upper limit of $3M_{\odot}$ for neutron stars.
In $f(R,T)$ scenario, Moraes et al. \cite{26} analyzed the features
of compact objects and determined that the higher values of coupling
parameter provide the structures of compact stars with larger
masses. Astashenok et al. \cite{41} investigated the properties of
neutron and quark stars in $f(R)$ gravity and found that positive
values of the model parameter yields stable solutions, whereas the
negative values lead to unstable stellar structures. They also
observed that the values of masses decrease with the increasing
values of the model parameter. In $f(R,T)$ framework, Sharif and
Siddiqa \cite{42} considered the MIT bag model and polytropic EoS to
examine the nature of quark stars and white dwarfs, respectively.
They observed that the masses of compact stars can cross the
Chandrasekhar and TOV limits in the presence of higher-curvature
terms.

It is worthwhile to mention here that the maximum values of masses
obtained in $f(R,T,Q)$ gravity lie very well within the required
limits in GR and larger values of the model parameter $\delta$ lead
to the smaller masses of compact stars. It is also noticed that this
gravity provides the existence of very small stellar objects as the
radii obtained in this gravity are smaller than the radii observed
in the above mentioned works.

\vspace{.25cm}

{\bf Acknowledgment}

\vspace{0.25cm}

We would like to thank the Higher Education Commission, Islamabad,
Pakistan for its financial support through the {\it Indigenous Ph.D.
5000 Fellowship Program Phase-II, Batch-III.}

\end{document}